\documentclass[prl,twocolumn,showpacs]{revtex4}
\usepackage{amssymb}

\begin{document}

\title{Catalytic Conversion Probabilities for Bipartite Pure States}
\author{S. Turgut}
\email{sturgut@metu.edu.tr}
\affiliation{ Department of Physics,
Middle East Technical University, 06531, ANKARA, TURKEY}


\begin{abstract}
For two given bipartite-entangled pure states, an expression is
obtained for the least upper bound of conversion probabilities
using catalysis. The attainability of the upper bound can also be
decided if that bound is less than one.
\end{abstract}



\maketitle

\section{Introduction}

A major problem in
quantum information theory is to understand the conditions for
transforming a given entangled state into another desired state by
using only local quantum operations assisted with classical
communication (LOCC). Significant development has been achieved
for the case of pure bipartite states. Bennett \emph{et al.} have
shown that for the asymptotic case, where essentially an infinite
number of copies of states are needed to be transformed,
conversion is possible as long as the entropy of entanglement is
conserved.\cite{Bennett1}

Away from the asymptotic limit, where a single copy of a given
state is to be transformed into another given state, such a simple
conversion criterion cannot be found and investigations have
unearthed a deep connection of the problem to the mathematical
theory of majorization.\cite{NielsenMaj} For setting up the
necessary notation, the following definitions are introduced
first. For two sequences with $n$ elements $x$ and $y$, we say
that $x$ is \emph{super-majorized} by $y$ (written $x\prec^w y$),
if $F_m(x)\ge F_m(y)$ for all $m=1,2,\ldots,n$. Here, $F_m(x)$
denotes the sum of the smallest $m$ elements of $x$, i.e.,
$F_m(x)=x_1^\uparrow+x_2^\uparrow+\cdots+x_m^\uparrow$, where
$x^\uparrow$ is the sequence $x$ with all elements arranged in
non-decreasing order ($x_1^\uparrow\le x_2^\uparrow\le\cdots\le
x_n^\uparrow$). If, in addition to these, the two sequences have
the same sum ($F_n(x)=F_n(y)$) then we say that $x$ is
\emph{majorized} by $y$ (written $x\prec y$).

Given two entangled states in Schmidt form,
$\vert\psi\rangle = \sum_{i=1}^n \sqrt{x_i}\vert i_A\otimes i_B\rangle$ and
$\vert\phi\rangle = \sum_{i=1}^n \sqrt{y_i} \vert i_A^\prime \otimes i_B^\prime\rangle$,
where $x$ and $y$ are the respective Schmidt coefficients ($\sum x_i=\sum y_i=1$), the
problem is essentially to determine the probability of converting
the state $\vert\psi\rangle$ into $\vert\phi\rangle$ by LOCC. As
two entangled states with the same Schmidt coefficients are
equivalent under local unitaries, that probability depends only on
the Schmidt coefficients and not on the particular local
orthonormal bases in which they are expressed. For that reason,
the conversion probability of $\vert\psi\rangle$ into
$\vert\phi\rangle$ will be simply denoted by $P(x\rightarrow y)$.

The most important step in the solution of this problem is taken
by Nielsen who has shown that $\vert\psi\rangle$ can be converted
into $\vert\phi\rangle$ with certainty, i.e., $P(x\rightarrow
y)=1$, if and only if $x\prec y$.\cite{Nielsen} Subsequently,
Vidal has obtained the expression $P(x\rightarrow y) = \min_{1\le
m \le n} F_m(x)/F_m(y)$ for the conversion probability between two
arbitrary states.\cite{Vidal} Note that the conversion probability
is equal to the largest value of $\lambda$ such that $x$ is
super-majorized by $\lambda y$, i.e.,
\begin{equation}
 P(x\rightarrow y)=\max \left\{ \lambda: \lambda\ge0,\quad x\prec^w \lambda y\right\}\quad,
\end{equation}
where $\lambda y$ denotes the sequence obtained by multiplying
each element of $y$ with $\lambda$.

An interesting development came with the demonstration of Jonathan
and Plenio that entangled pairs can be used just like catalysts to
improve conversion probabilities.\cite{JonathanPlenio} To be explicit, if
$\vert\chi\rangle=\sum_{\ell=1}^N \sqrt{c_\ell}\vert
\ell_A\otimes\ell_B\rangle$ is another entangled state shared by
the same parties, then for some cases
$\vert\psi\rangle\otimes\vert\chi\rangle$ can be converted into
$\vert\phi\rangle\otimes\vert\chi\rangle$ with a probability more
than that of $\vert\psi\rangle$ to $\vert\phi\rangle$ conversion.
In terms of the Schmidt coefficients we have $P(x\otimes
c\rightarrow y\otimes c)\ge P(x\rightarrow y)$, where strict
inequality is obtained for some cases. In such a transformation,
the entangled state $\vert\chi\rangle$ is not consumed, although
it takes part in the transformation much like a catalyst in
chemical reactions.

Subsequently, a lot of research has been directed to understanding
the catalytic transformations.\cite{Feng1,Daftuar,Feng2} A major
problem to be solved is to determine the catalytic conversion
probability, i.e., $P_{\text{cat}}(x\rightarrow y)=\sup_c
P(x\otimes c\rightarrow y\otimes c)$, where the supremum is taken
over all finite sequences $c$ of positive numbers. This quantity
is actually the least upper bound on catalytic conversion
probabilities as it may not be possible to attain the probability
value $P_{\text{cat}}(x\rightarrow y)$ by a reasonable catalyst
$c$. However, for any probability smaller than the bound,
catalysis is possible.

Nielsen has suggested the term $x\prec_T y$ ($x$ is \emph{trumped}
by $y$) whenever there is a $c$ such that $x\otimes c\prec
y\otimes c$.\cite{NielsenMaj} The notation will be extended and we
will say that $x$ is \emph{super-trumped} by $y$ (written
$x\prec_T^wy$) if there is a $c$ such that $x\otimes c\prec^w
y\otimes c$. The catalytic conversion probability can be expressed
with this notation as
\begin{equation}
P_{\text{cat}}(x\rightarrow y)=\sup\left\{ \lambda: \lambda\ge0,
~x\prec_T^w\lambda y\right\}\quad.
\end{equation}
The purpose of this letter is to provide a computable expression
for that probability, mainly by finding all of the necessary and
sufficient conditions for $x\prec_T^wy$ relation for the case
$\sum x_i>\sum y_i$. As the case $\sum x_i=\sum y_i$ is not covered, the
results in this letter will not enable us to analyze the trumping
relation.

First, let us define $\nu$th power mean of an $n$-element sequence
$x$ as
\begin{equation}
  A_\nu(x)=\left(\frac{1}{n}\sum_{i=1}^n x_i^\nu\right)^\frac{1}{\nu}\quad.
\end{equation}
For all finite $\nu$, this is a continuous function which has a
limit $A_{-\infty}(x)=x_1^\uparrow$. For the particular value
$\nu=0$, it gives the geometric mean $A_0(x)=(\prod x_i)^{1/n}$.
Note that, if any element of the sequence $x$ is zero, then
$A_\nu(x)=0$ for all $\nu\le0$. We would like to prove the following theorem.

\textbf{Theorem:} If $x$ and $y$ are $n$-element sequences of
non-negative numbers such that $x$ has only positive elements and
$\sum x_i>\sum y_i$, then $x\prec_T^w y$ if and only if
\begin{equation}
  A_\nu(x)>A_\nu(y)\quad,\quad\forall~\nu\in(-\infty,1)\quad.
\label{eq:Conditions}
\end{equation}
Note that the inequalities are strict and the end point
$\nu=-\infty$ is not included ($\nu=1$ is also strict by
assumption).

Even though the theorem deals only with the special case
$\sum x_i>\sum y_i$, it is nevertheless possible to express the
catalytic transformation probability as
\begin{equation}
P_{\text{cat}}(x\rightarrow y) = \min_{\nu\in[-\infty,1]}\frac{A_\nu(x)}{A_\nu(y)}\quad.
\label{eq:CatalyticConversionProbability}
\end{equation}
where $\min$ is used by the inclusion of the end points. Although
the minimization is over a continuous variable, it is possible to
compute $P_{\text{cat}}(x\rightarrow y)$ to any desired accuracy.
Moreover, the theorem tells us that if there is a $\nu$ in the
interval $(-\infty,1)$ that attains the minimum of
(\ref{eq:CatalyticConversionProbability}), then
$P_{\text{cat}}(x\rightarrow y)$ can not be achieved by any
catalyst $c$ (e.g., when $P_{\text{cat}}(x\rightarrow
y)<\min(1,x_1^\uparrow/y_1^\uparrow)$). On the other hand, if this
is not the case and $P_{\text{cat}}(x\rightarrow y)<1$, then that
value can be achieved by some catalyst $c$.

The following facts, which are not too difficult to prove, will be
frequently used. (1) For any sequence $x$, we define the
\emph{characteristic function} $H_x(t) = \sum_{i=1}^n (t-x_i)^+$
where $(\alpha)^+=\max(\alpha,0)$ denotes the positive-part
function. Super-majorization relation $x\prec^w y$ between two
non-negative sequences can be equivalently stated\cite{NielsenMaj}
as $H_x(t) \le H_y(t)$ for all $t\ge0$. (2) Moreover, if
$x^\uparrow\neq y^\uparrow$, then $H_y(t)-H_x(t)$ is strictly
positive on some interval. (3) For the cross-product of two
sequences we have $H_{x\otimes c} =\sum_\ell c_\ell
H_x(t/c_\ell)$. (4) If all elements of $\bar{x}$ is greater than
the corresponding elements of $x$, i.e., $\bar{x}_i\ge x_i$, then
$\bar{x}\prec^wx$. (5) If  $x\prec^w y$ then $x\prec_T^w y$. (6)
Finally, $\prec^w$ and $\prec_T^w$ are partial orders on sequences
with $n$-elements.

\textit{Proof of necessity:} It will be shown that if $x\prec_T^w
y$, $x$ has no zero elements and $x^\uparrow\neq y^\uparrow$, then
the inequalities (\ref{eq:Conditions}) are satisfied (it is not
necessary to assume $\sum x_i>\sum y_i$). There is a sequence $c$
having positive elements such that $\Delta(t) = H_{y\otimes
c}(t)-H_{x\otimes c}(t)$ is non-negative.
For $t>c_{\max}\max(x_n^\uparrow,y_n^\uparrow)$,
the function $\Delta(t)$ has the constant value $(\sum x_i-\sum y_i)\sum c_\ell$. For that reason, the integral
\begin{equation}
  I_\nu = \int_0^\infty \Delta(t) t^{\nu-2} dt
\end{equation}
is convergent at $t=\infty$ for all values of $\nu<1$. Moreover,
(i) if $y$ has no zero elements, then $\Delta(t)=0$ for a
sufficiently small $t$ and the integral is convergent at $t=0$.
(ii) If $y$ has zero entries, then $\Delta(t)\propto t$ near $t=0$
and therefore the integral is convergent only for $0<\nu<1$; but
this is sufficient for us as (\ref{eq:Conditions}) is satisfied
for all $\nu\le0$. Finally, strict positivity of $\Delta(t)$ in
some interval implies that $I_\nu$ is strictly positive. Since the
integral is
\begin{equation}
I_\nu=\left\{
\begin{array}{ll}
\frac{1}{\nu(1-\nu)} \left(\sum_{i=1}^n x_j^\nu-y_j^\nu\right)\sum_\ell c_\ell^\nu & \nu\neq0 \\
\left(\ln\prod x_i/\prod y_i\right)\left(\sum_\ell 1\right) &
\nu=0
\end{array}
\right.
\end{equation}
investigating $\nu<0$, $\nu=0$ and $\nu>0$ cases separately, it
can be seen that (\ref{eq:Conditions}) are satisfied.$\Box$

Proof of sufficiency is lengthy and needs the introduction of a
separate problem. Let $\gamma(s)=\sum_{m=0}^N \gamma_m s^m$ be a
real polynomial where some of the coefficients $\gamma_m$ might be
negative. The problem is to express $\gamma$ as a ratio of two
power series with non-negative coefficients, which are required to
be convergent at a desired value $s=R$. To be precise, we would
like to find two power series $a(s)=\sum_{m=0}^\infty a_m s^m$ and
$b(s)=\sum_{m=0}^\infty b_m s^m$ such that (i)
$a(s)\gamma(s)=b(s)$, (ii) $a_m\ge 0$ and $b_m\ge0$ for all $m$
and finally (iii) both $a(R)$ and $b(R)$ are finite. We will say
that $\gamma$ belongs to the polynomial set $\mathcal{P}_R$ when
this problem has a solution. It is obvious that if
$\gamma\in\mathcal{P}_R$, then $\gamma(s)>0$ for all $s\in(0,R]$.
The following lemma shows that this property is also sufficient.

\textbf{Lemma:} For a polynomial $\gamma(s)$, if $\gamma(s)>0$ for
all $s$ in the range $0<s\le R$ then $\gamma\in\mathcal{P}_R$.

\textit{Proof:} First, note that the product of two elements of
$\mathcal{P}_R$ is in the same set. For if
$\gamma_1,\gamma_2\in\mathcal{P}_R$ and $a_i$ and $b_i$ are the
respective series satisfying positive coefficient and convergence
properties such that $a_i(s)\gamma_i(s)=b_i(s)$ for $i=1,2$, then
we have $a_1(s)a_2(s)\gamma_1(s)\gamma_2(s)=b_1(s)b_2(s)$. Since
$a_1a_2$ and $b_1b_2$ are convergent at $R$ and have non-negative
series coefficients we have $\gamma_1\gamma_2\in\mathcal{P}_R$.
For that reason, the assertion will first be proven for
irreducible factors of $\gamma$.

(1) For $\gamma(s)=1-\xi s$, it will be shown that if $\xi R<1$
then $\gamma\in\mathcal{P}_R$. For the case, $\xi\le0$, there is
nothing to be shown as $\gamma$ has already non-negative
coefficients. For the case, $0<\xi R<1$, we have $a(s)=(1-\xi
s)^{-1}=\sum_{m=0}^\infty \xi^m s^m$ and $b(s)=1$, which satisfy
the requirements, so that we have $\gamma\in\mathcal{P}_R$.

(2) For $\gamma(s)=1-2\xi s+\lambda s^2$, it will be shown that if
$\lambda>\xi^2$ then $\gamma\in\mathcal{P}_R$. Obviously, for
$\xi\le0$ there is nothing to be proven, so consider $\xi>0$ for
the following. Let $N$ be an integer sufficiently large so that
\begin{equation}
  \frac{1}{4} \left(\frac{(2N)!}{N!^2}\right)^\frac{1}{N} \ge \frac{\xi^2}{\lambda}\quad.
\end{equation}
We can always find such an $N$ as the left-hand side has limit $1$
as $N\rightarrow\infty$ and the right-hand side is strictly less
than $1$. In that case we choose
\begin{eqnarray}
a(s) &=& \sum_{k=0}^{2N-1} (1+\lambda s^2)^k (2\xi s)^{2N-1-k}\quad,\\
b(s) &=& (1+\lambda s^2)^{2N}-(2\xi s)^{2N}\quad.
\end{eqnarray}
Note that all coefficients of $a$ are already non-negative. That is true for $b$ as
well, since the coefficient of $s^{2N}$ is
$\lambda^{N}(2N)!/N!^2 -(2\xi)^{2N}$
which is also non-negative by the special choice of $N$. Therefore, $\gamma\in\mathcal{P}_R$.

(3) The lemma can now be proven for a general polynomial. Express
$\gamma(s)$ as a product of its irreducible factors as
\begin{equation}
  \gamma(s)= A s^r \prod_i (1-\xi_i s)\prod_i(1-2\xi_i^\prime s+(\xi_i^{\prime2}+\eta_i^{\prime2})s^2)\quad,
\end{equation}
where $r(\ge0)$ is the multiplicity of a possible root at $0$,
$1/\xi_i$ are the real roots, $(\xi_i^\prime\pm
i\eta_i^\prime)^{-1}$ are the complex roots of $\gamma$ and $A>0$.
Since $\gamma$ is non-zero on the interval $(0,R]$, each real root
satisfies $\xi_iR<1$. As each factor is in $\mathcal{P}_R$, we have
$\gamma\in\mathcal{P}_R$.$\Box$

Note that if $R>1$ and $\gamma\in\mathcal{P}_R$, then the infinite
series $a(s)$ can be chosen such that the value $a(1)$ and all
series coefficients $a_m$ are rational numbers. The reason is that
$a(s)$ and $b(s)$ can both be multiplied by a third series which
satisfies the necessary non-negativity and convergence properties.
By choosing the coefficients of the third series, all of these
numbers can be made rational simultaneously as the reader can
easily check. After this brief diversion, we can continue with the
rest of the proof of the theorem.

\textit{Proof of sufficiency:} If two $n$-element sequences $x$
and $y$ (such that $x^\uparrow\neq y^\uparrow$) share some common
elements, then the corresponding elements can be removed from
each, which gives shorter sequences $\bar{x}$ and $\bar{y}$ (which
have no common elements, i.e., $\bar{x}_i\neq\bar{y}_j$). It is
easy to verify that (i) $x\prec^w y$ iff $\bar{x}\prec^w \bar{y}$,
(ii) $x\prec_T^w y$ iff $\bar{x}\prec_T^w\bar{y}$, and (iii)
$A_\nu(x)>A_\nu(y)$ iff $A_\nu(\bar{x})>A_\nu(\bar{y})$. For this
reason, it is only necessary to give the proof for sequences which
have no common elements. This will be assumed below. It will also
be assumed that $x$ and $y$ are arranged in non-decreasing order
($x=x^\uparrow$ and $y=y^\uparrow$). The complete proof of the
sufficiency of the inequalities (\ref{eq:Conditions}) will be
completed in three steps, each one being in the form of a separate
theorem dealing with a special case.

\textbf{Case A.} $y$ has strictly positive elements such that
$y_i=K\omega^{\alpha_i}$ and $x_i=K\omega^{\beta_i}$ for some
integers $\alpha_i$ and $\beta_i$ and for some numbers $K>0$ and
$\omega>1$.

\textit{Proof:} Redefine $K$ such that $\alpha_1=0$ (as a result,
$\alpha_i\ge0$ for all $i$) and then set $K=1$ by dividing each
sequence by a common number. Note that $\nu\rightarrow-\infty$
limit of (\ref{eq:Conditions}) gives $x_1\ge y_1$. As $x$ and $y$
have no common elements, we have $\beta_i>0$ for all $i$. Let the
polynomial $\Gamma(s)$ be defined as
\begin{equation}
  \Gamma(s)=\sum_{i=1}^n (s^{\alpha_i}-s^{\beta_i})=\sum_{k}\Gamma_k s^k\quad,
\end{equation}
and let $\gamma(s)=\Gamma(s)/(1-s)$. Since $\Gamma(1)=0$,
$\gamma(s)$ is also a polynomial. We will first show that
$\gamma\in\mathcal{P}_\omega$. The inequality
(\ref{eq:Conditions}) at $\nu=0$ implies that
$\gamma(1)=\sum_{i=1}^n(\beta_i-\alpha_i)$ is strictly positive.
Next, let $s=\omega^\nu$ where $\nu$ is any value in $(-\infty,1]$
excluding $\nu=0$. In that case, we have
\begin{eqnarray}
  \gamma(s) &=& \frac{1}{1-\omega^\nu}\sum_{i=1}^n  (y_i^\nu-x_i^\nu)\quad.
\end{eqnarray}
Investigating the cases $\nu<0$ and $\nu>0$ separately, one finds
that $\gamma(s)>0$. As a result, we have
$\gamma\in\mathcal{P}_\omega$.

By the lemma, there exists two (possibly infinite) series $a(s)$
and $b(s)$ which are convergent at $s=\omega$ and have
non-negative series coefficients. Moreover, $a(s)$ will be chosen
in such a way that all of its coefficients and $a(1)$ are rational
numbers. As $\gamma(0)>0$, $a_0$ and $b_0$ can be made non-zero.
The relationship $a(s)\Gamma(s)=(1-s)b(s)$ implies that
$\sum_{k=0}^m a_k \Gamma_{m-k} = b_m-b_{m-1}$, where we define
$b_{-1}=0$ for simplicity.

Let $\bar{h}(t)= \sum_{m=0}^\infty a_m (t-\omega^m)^+$, a function
which is a sum of a finite number of terms for any fixed $t$. Let
\begin{eqnarray}
 \bar{\delta}(t) &=& \sum_{i=1}^n
 y_i\bar{h}\left(\frac{t}{y_i}\right)-x_i\bar{h}\left(\frac{t}{x_i}\right)   
    = \sum_k \Gamma_k \omega^k \bar{h}(t\omega^{-k})\quad,\nonumber\\
    &=& \sum_{m=0}^\infty (b_m-b_{m-1})(t-\omega^m)^+\quad.
\end{eqnarray}
It can be shown that $\bar{\delta}(t)\ge0$ for all $t\ge0$, but
better lower bounds can be placed as follows: (i) For $t\le\omega$, we have
$\bar{\delta}(t)=b_0(t-1)^+\ge0$. (ii) For $t\ge\omega$, there is an integer $N\ge1$ such that
$\omega^N\le t\le \omega^{N+1}$ and we have
\begin{equation}
\bar{\delta}(t)=b_N(t-\omega^N)+(\omega-1)\sum_{m=0}^{N-1}
b_m\omega^m\ge (\omega-1)b_0
\end{equation}
i.e., a strictly positive lower bound.


Let $\epsilon=(\omega-1)b_0/(\sum_k\vert\Gamma_k\vert\omega^k)$.
Since $a(\omega)<\infty$, we can find an integer $M(\ge1)$ such
that $\sum_{m=M}^\infty a_m \omega^m <\epsilon/2$. Define
$A=\sum_{m=M}^\infty a_m$. This is a rational number and
satisfies the inequality $A\omega^M<\epsilon/2$. Consider the
function
\begin{equation}
  h(t)=\sum_{m=0}^{M-1} a_m (t-\omega^m)^+  +   A(t-\omega^M)^+\quad.
\end{equation}
The following bounds can be placed on $\vert\bar{h}(t)-h(t)\vert$:
(i) If $t\le\omega^M$ we have $h(t)=\bar{h}(t)$. (ii) If
$t\ge\omega^M$, there is an $N\ge M$ such that $\omega^N\le t\le
\omega^{N+1}$ and
\begin{eqnarray}
  \left\vert\bar{h}(t)-h(t)\right\vert
   &=& \left\vert  A\omega^M -\sum_{m=M}^N a_m \omega^m-\sum_{m=N+1}^\infty a_m t\right\vert \nonumber\\
   &\le& A\omega^M+\sum_{m=M}^\infty a_m \omega^m <\epsilon\quad.
\end{eqnarray}
As a result, the following function
\begin{eqnarray}
  \delta(t) &=& \sum_{i=1}^n y_ih\left(\frac{t}{y_i}\right)-x_ih\left(\frac{t}{x_i}\right) 
    = \sum_k \Gamma_k \omega^k h(t\omega^{-k})\quad, \nonumber \\
    &=& \bar{\delta}(t)+\sum_k \Gamma_k \omega^k (h(t\omega^{-k})-\bar{h}(t\omega^{-k}))
\end{eqnarray}
is non-negative everywhere since (i) for $t\le\omega$
we have $\delta(t)=\bar{\delta}(t)\ge0$ and (ii) for
$t\ge\omega$ we have
$\delta(t)>\bar{\delta}(t)-\sum_{k}\vert\Gamma_k\vert\omega^k\epsilon=\bar{\delta}(t)-(\omega-1)b_0\ge0$.

Let $\mathcal{N}$ be a sufficiently large integer so that all of
$\mathcal{N}a_0,\mathcal{N}a_1,\ldots,\mathcal{N}a_{M-1},\mathcal{N}A$
are integers. Schmidt coefficients of the catalyst sequence $c$
will be chosen as $\omega^m$, repeated $\mathcal{N}a_m$ times (for
$0\le m\le M-1$), and as $\omega^M$, repeated $\mathcal{N}A$
times. Then $H_c(t)=\mathcal{N}h(t)$ is the characteristic
function of $c$ and the non-negativity of $\delta(t)$ is
equivalent to $x\otimes c\prec^w y\otimes c$. This proves our
assertion that $x\prec_T^w y$.$\Box$

\textbf{Case B.} $y$ has strictly positive elements.

\textit{Proof:} As $x$ and $y$ have no common elements, the
inequalities (\ref{eq:Conditions}) imply that
$x_1^\uparrow>y_1^\uparrow$. Let, $\theta=\min_{\nu\in[-\infty,1]}
A_\nu(x)/A_\nu(y)$. Since the end points are included, the minimum
exists and therefore $\theta>1$. Let $\omega=\theta^{1/3}$ and
define two $n$-element sequences $\bar{x}$ and $\bar{y}$ as
$\bar{y}_i=\omega^{\alpha_i}$ and $\bar{x}_i=\omega^{\beta_i}$
where
\begin{equation}
  \alpha_i = \left]\frac{\ln y_i}{\ln\omega}\right[ \quad,\quad
  \beta_i = \left[\frac{\ln x_i}{\ln\omega}\right] \quad,
\end{equation}
$[t]$ is the largest integer smaller than $t$ and $]t[$ is the
smallest integer greater than $t$. Using ${]t[}-1<t\le]t[$ and $[t]\le t<[t]+1$, we get
\begin{equation}
  \frac{\bar{y}_i}{\omega}<y_i\le \bar{y}_i\quad,\quad
  \bar{x}_i \le x_i < \omega\bar{x_i}\quad.
\label{eq:xyandbarinequalities}
\end{equation}
Then for any $\nu\in[-\infty,1]$ we have
\begin{equation}
A_\nu(\bar{x})>\frac{1}{\omega}A_\nu(x)\ge\frac{\theta}{\omega}A_\nu(y)>\frac{\theta}{\omega^2}A_\nu(\bar{y})\quad.
\end{equation}
As a result, $A_\nu(\bar{x})>A_\nu(\bar{y})$ for all
$\nu\in[-\infty,1]$; $\bar{x}$ and $\bar{y}$ fulfills the
conditions of case A, and therefore $\bar{x}\prec_T^w\bar{y}$.
Finally, the inequalities (\ref{eq:xyandbarinequalities}) imply
$x\prec^w\bar{x}$ and $\bar{y}\prec^wy$. All of these prove our
assertion that $x\prec_T^w y$.$\Box$

\textbf{Case C.} $y$ has zero elements.

The proof will be carried out by replacing all zero elements of
$y$ with a small value $\epsilon$ in such a way that this case is
reduced to case B. Suppose that $y$ has exactly $m$ entries equal
to 0 ($0<m<n$). Note that the inequalities (\ref{eq:Conditions})
are automatically satisfied for $\nu\le0$. Using the premise that
(\ref{eq:Conditions}) are satisfied for $\nu\in(0,1]$, we can
deduce that the function
\begin{equation}
  J_\nu =\left(\frac{\sum_{i=1}^n x_i^\nu - \sum_{i=m+1}^n y_i^\nu}{m}\right)^\frac{1}{\nu}  \quad,
\end{equation}
is strictly positive for all $\nu\in(0,1]$. Moreover, it has a
positive limit $J_0 = \left(\prod x/\prod_{i=m+1}^n
y_i\right)^{1/m}$ at the end point $\nu=0$. As a result,
$J_{\min}=\min_{\nu\in[0,1]} J_\nu$ exists and is non-zero as the
minimum is taken over a compact interval. Let $\epsilon$ be a
positive number such that
\begin{equation}
\epsilon<\min\left(J_{\min},y_n\left(\frac{x_1}{y_n}\right)^\frac{n}{m}\right)\quad,
\end{equation}
and define a new sequence $\bar{y}$ as
$\bar{y}_1=\cdots=\bar{y}_m=\epsilon$ and $\bar{y}_i=y_i$ for all
$i>m$. It is obvious that $\bar{y}\prec^w y$. Showing that
$x\prec_T^w\bar{y}$ will complete the proof. For this purpose, we
look at the power means. (i) For $\nu\in(0,1]$, it is trivial to
check that $J_\nu >\epsilon$ is equivalent to
$A_\nu(x)>A_\nu(\bar{y})$. (ii) For $\nu=0$ we have
\begin{eqnarray}
\frac{A_0(x)}{A_0(\bar{y})} =
  \frac{\left(\prod_{i=1}^n x_i \right)^\frac{1}{n}}%
{\left(\epsilon^m \prod_{i=m+1}^n y_i\right)^\frac{1}{n}} \ge
\frac{x_1}{y_n}\left(\frac{y_n}{\epsilon}\right)^\frac{m}{n}>1~.
\end{eqnarray}
(iii) For $\nu<0$, we use Bernoulli's inequality, which states
that $\alpha^r-1\ge r(\alpha-1)$ for all $r\ge1$ and $\alpha>0$,
as follows
\begin{eqnarray}
  m(\epsilon^\nu -y_n^\nu) &>& m  y_n^\nu\left(\left(\frac{x_1}{y_n}\right)^{\nu\frac{n}{m}}-1\right)
        \\  
        &\ge& m  y_n^\nu\frac{n}{m} \left(\left(\frac{x_1}{y_n}\right)^\nu-1\right) \\
        &=& n(x_1^\nu-y_n^\nu) \quad,
\end{eqnarray}
which implies that
\begin{eqnarray}
  \sum_{i=1}^n \bar{y}_i^\nu &=& m\epsilon^\nu+\sum_{i=m+1}^n y_i^\nu 
    \ge m\epsilon^\nu+(n-m)y_n^\nu \\
    &>& nx_1^\nu \ge \sum_{i=1}^n x_i^\nu
\end{eqnarray}
The result $A_\nu(x)>A_\nu(\bar{y})$ follows from here. As power
mean inequalities are satisfied for all $\nu\in(-\infty,1]$, we
have $x\prec_T^w\bar{y}$ by the result in case B, which completes
the proof. $\Box$


\end{document}